\documentclass[10pt]{IEEEtran}

\ifCLASSINFOpdf

\fi

\usepackage[subrefformat=parens,labelformat=parens,caption=false,font=footnotesize]{subfig}
\usepackage{array}
\usepackage{balance}
\usepackage{amsmath}
\usepackage{breqn}
\usepackage[nolist,nohyperlinks]{acronym}
\usepackage{wrapfig,lipsum}
\usepackage{graphicx}
\usepackage{epstopdf}
\usepackage{rotating}
\usepackage{amsmath, amssymb, amsthm}
\usepackage{stmaryrd}
\usepackage{tikz}
\usepackage{hyperref}
\usepackage{verbatim}
\usepackage{url}
\usepackage[utf8]{inputenc}
\usepackage[english]{babel}

\usepackage{multicol}
\usepackage{xr-hyper}

\usepackage{scalerel}
\usepackage{multicol}

\usepackage[noadjust]{cite}
\usepackage[normalem]{ulem}
\usepackage{color}
\usepackage{dsfont}
\usepackage{bm}
\usepackage[short]{optidef} 
\usepackage{diagbox}
\usepackage{enumitem}
\usepackage{slashbox}
\usepackage[ruled]{algorithm2e}
\usepackage{multirow}
\usepackage[T1]{}
\usepackage{color, colortbl}
\usepackage{babel}
\usepackage{verbatim}
\usepackage{textcomp}
\usepackage{tabulary}
\usepackage{booktabs}
\usepackage{caption}
\definecolor{Gray}{gray}{0.95}
\definecolor{LightCyan}{rgb}{0.8,0.85,1}
\definecolor{LightBlue}{rgb}{0.6,0.6,1}
\usepackage[font=small]{caption}
\usepackage{mathtools}

\setlist{nosep}

\usepackage{mathtools}

\begin{document}

\title{Large Language Models for Telecom: \\
Forthcoming Impact on the Industry}
\author{Ali Maatouk$^*$, Nicola Piovesan$^*$, Fadhel Ayed$^*$, Antonio De Domenico$^*$, and Merouane Debbah$^\dagger$\\
$^*$Paris Research Center, Huawei Technologies, Boulogne-Billancourt, France
\\ $^\dagger$Khalifa University of Science and Technology, Abu Dhabi, UAE}
\maketitle
\thispagestyle{empty}

\begin{abstract}
Large Language Models (LLMs)\textcolor{black}{, AI-driven models that can achieve general-purpose language understanding and generation,} have emerged as a transformative force, revolutionizing fields well beyond Natural Language Processing (NLP) and garnering unprecedented attention. As LLM technology continues to progress, the telecom industry is facing the prospect of its impact on its landscape. To elucidate these 
implications, we delve into the inner workings of LLMs, providing insights into their current capabilities and limitations. We also examine the use cases that can be readily implemented in the telecom industry, streamlining tasks\textcolor{black}{, such as anomalies resolutions and technical specifications comprehension, which} currently hinder operational efficiency and demand significant manpower and expertise. Furthermore, we uncover essential research directions that deal with the distinctive challenges of utilizing the LLMs within the telecom domain. Addressing them represents a significant stride towards fully harnessing the potential of LLMs and unlocking their capabilities to the fullest extent within the telecom domain.
\end{abstract}

\section{Introduction}
\label{sec:intro}
\acp{LLM} have revolutionized \ac{NLP} and \ac{AI}, propelling text generation, comprehension, and interaction to \textcolor{black}{unprecedented levels of sophistication.}
The history of \acp{LLM} can be traced back to the early developments in \ac{ML} and \ac{NLP}, which encompassed the emergence of statistical language models and the advancements in neural networks. However, it was the rise of transformer architectures \cite{NIPS2017_3f5ee243}, which paved the way for the development of language models capable of processing and generating vast amounts of text. Among the notable advancements in this domain, OpenAI's \ac{GPT} series \textcolor{black}{and open-source LLMs like LLaMA and its successor LLaMA2 have} garnered significant attention \cite{openai2023gpt4}. 
Specifically, they have surpassed earlier models in terms of scale and capability, empowering human-like language understanding and generation.

\textcolor{black}{Thanks to their language understanding capabilities, \acp{LLM} have the potential to revolutionize diverse domains~\cite{Hadi2023}}, surpassing traditional \ac{NLP} applications like machine translation and sentiment analysis. In fact, through domain-specific data, they can excel in tasks related to that particular domain. For instance, in medicine, \acp{LLM} may play a crucial role in encoding clinical knowledge and supporting medical decision-making processes. 
Similarly, researchers in finance have investigated how \acp{LLM} can provide insights into market trends and assist in risk analysis. \textcolor{black}{Also, educational organizations have recently developed an LLM-based virtual tutor
and classroom assistant.}

Although \acp{LLM} have already demonstrated their potential in various fields, their application in the telecom industry has been relatively scarce. 
However, this situation is changing as more researchers are beginning to explore the capabilities of \acp{LLM} in this domain. For instance, a \ac{BERT}-like language model was adapted to the telecom domain \cite{holm2021bidirectional} to test its ability to answer a small, manually curated dataset of telecom questions. In another work, \textcolor{black}{language models} such as \ac{BERT} and \ac{GPT}-2 were leveraged to classify working groups within the \ac{3GPP} based on analysis of technical specifications \cite{bariah2023understanding}. Moreover, the potential of \acp{LLM} in facilitating Field-Programmable Gate Array development within wireless systems was highlighted in \cite{du2023power}. 
Additionally, the authors in \cite{bariah2023large} provided a vision where \acp{LLM}, \color{black}along with multi-modal data (e.g., images)\color{black}, can significantly contribute to the development of \ac{RAN} technologies such as beamforming and localization. In this future, by combining different data types like text and visuals, \acp{LLM} can assist in optimizing and improving \ac{RAN} functionalities. 

\textcolor{black}{
In parallel to the work initiated by the research community, telecom ecosystem industries offer the first products based on \ac{LLM} technologies. Huawei has released Pangu, an \ac{LLM} that has been tested in mining, government, vehicles, weather, and R$\&$D applications. Qualcomm has released an AI engine to support up to 10 billion parameters of generative \ac{AI} models on mobile handsets, allowing \ac{AI} assistant with \ac{NLP} capabilities and image generations based on Stable Diffusion. Moreover, Google has introduced generative \ac{AI} capabilities in its cloud platform to offer \acp{MNO} the opportunity to integrate \ac{NLP} functionalities in applications such as root cause analysis, information retrieval in legal documents, and conversational chatbot for customer experience improvement.}

In light of these applications, a fundamental question arises regarding the immediate and future impact of \acp{LLM} on the telecom industry. In this article, we aim to answer this question by providing a view of \acp{LLM} and their impeding influence on the industry. Our objective is to demystify their current abilities, highlight their existing limitations, and showcase several use cases in the telecom industry where they can provide substantial assistance today. Additionally, we highlight the telecom data within the industry that can be harnessed to leverage the capabilities of \acp{LLM}. Moreover, we shed light on the technical difficulties that arise in implementing these use cases and outline the research directions that need to be pursued to fully harness the potential of \acp{LLM}.


\section{Demystifying \acp{LLM}}
\label{sec:demystifying}
To explore the potential of \acp{LLM} in the telecom industry, it is essential to begin by gaining an understanding of their intrinsic behavior. To do so, we delve into the intricacies of \acp{LLM} architecture and training, exploring their capabilities as well as their limitations.

\subsection{Fundamentals of \acp{LLM}}
\begin{figure}
    \centering
    \includegraphics[width=0.45\textwidth]{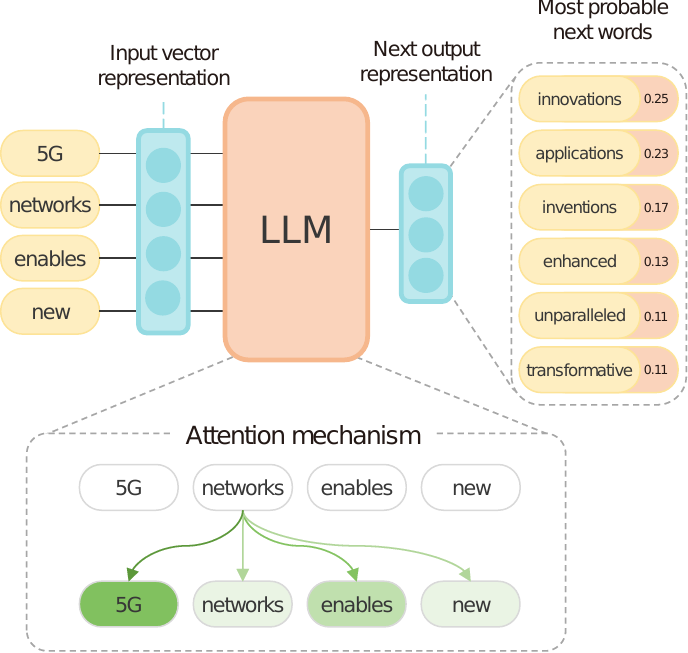}
    \caption{
    A high-level overview of \acp{LLM}.}
    \label{fig:examplearchitechture}
\end{figure}
\acp{LLM} are \ac{DL} models with the ability to process information and demonstrate human-like text generation capabilities.
Typically, \acp{LLM} utilize transformer-based architectures, where self-attention plays a pivotal role \cite{NIPS2017_3f5ee243}. In self-attention, each word in an input sequence attends to all the other words, calculating attention scores that signify the importance of each word relative to the others. This mechanism allows to effectively capture long-range dependencies and grasp the contextual usage of each word. \textcolor{black}{Interested readers can refer to the paper in~\cite{NIPS2017_3f5ee243} for
a mathematical description of the self-attention mechanism.}
In Fig \ref{fig:examplearchitechture}, a high-level illustration of an \ac{LLM} is presented, along with the accompanying self-attention mechanism. Another essential component in the transformer architecture is multi-head attention, which expands upon the concept of self-attention. Often, a sequence element needs to attend to multiple distinct aspects, and relying on a single attention mechanism alone is inadequate to accomplish this objective. The multi-head attention provides the flexibility by enabling the model to attend to different aspects of the input, capturing diverse patterns and dependencies within the input sequence. This capability allows the model to learn complex interactions between words and comprehensively understand the input. 

In addition, \acp{LLM} undergo extensive pretraining on vast amounts of text to acquire an understanding of the statistical properties inherent in the language at hand. \color{black}During this phase, the models are mainly trained with data crawled from the internet, which provides them with diverse
linguistic information. \color{black}The primary goal of this pretraining is to enable the model to predict the next word in a sentence based on the preceding words. 
Through this process, the model captures both syntactic and semantic relationships, thereby enhancing its grasp of contextual nuances. Due to the range of corpora used during training and the large number of model parameters involved, \acp{LLM} can develop a comprehensive understanding of grammar, reasoning abilities, and even comprehend intricate language structures.

Although the pretrained \ac{LLM} has a comprehensive understanding of the statistical properties within the language, it needs specific domain knowledge to be applied to industrial processes.
To achieve this, the pretrained \ac{LLM}'s parameters, including attention blocks, are fine-tuned using domain-specific datasets and similar training techniques employed during the pretraining phase. Through this procedure, referred to as knowledge fine-tuning, the \ac{LLM} can adapt the learned representations, denoted to as embeddings, from the pretraining phase to better align with the intricacies of the specific domain.
\textcolor{black}{In addition, researchers have designed prompt engineering solutions, such as chain-of-thought (CoT) prompting and \ac{RAG}, to enhance the capability of LLMs on a wide range of tasks. This topic and related open challenges are further discussed in Sec. \ref{sec:Foundation Model}.}

\subsection{\acp{LLM} Functionalities}
{\color{black}
The LLM's potential shines through its three core competencies: an extensive understanding of the intricacies of language, cross-disciplinary knowledge, and the emerging ability to reason, albeit less developed than the former two. While we discuss three distinct functionalities: semantic comprehension, intelligent knowledge retrieval, and orchestration capabilities, it is important to note their inherent overlap in practical applications, as highlighted in Section III.

\subsubsection{Semantic abilities} 
{\color{black} LLMs develop an internal representation of textual data in the form of real-valued vectors called embeddings. This representation conveniently encapsulates the input text's semantics, syntax, and contextual interpretation.} These embeddings provide a simplified representation of textual data suitable for algorithmic procedures and data analysis. For example, a large city's telecom network generates millions of daily trouble tickets. Many disruptions are symptomatic of the same core issues; however, due to compartmentalization within the network infrastructure, there is no automated system for categorizing these tickets. By converting them into embeddings from a domain-specific \ac{LLM}, clustering algorithms like K-Means can effectively group the tickets, potentially tying them back to singular faults.

\subsubsection{Intelligent access to knowledge} By understanding the specific intention conveyed through the prompt, an \ac{LLM} can effectively apply its knowledge base to craft a response tailored to the user's needs. \acp{LLM} can process and comprehend intricate information, such as the content within standard documents, discern patterns, and infer logical conclusions from the given inputs. \acp{LLM} thus transition from passive language processors to active and intelligent agents, functioning as assistants or co-pilots that enhance professionals' productivity. For instance, in an operations and maintenance scenario, an operator faced with a trouble ticket may benefit from the model's ability to summarize the issue automatically, suggest possible solutions, and even draft a template email for field engineers to act upon, requiring only the operator's review and approval.

\subsubsection{LLMs as orchestrators} LLMs can utilize their reasoning to deconstruct complex tasks into manageable subtasks and deploy suitable (external) tools for each. They manage workflows by identifying the most appropriate tool for each segmented operation. Take, for instance, a task such as forecasting the next day's energy consumption for a \ac{BS} undergoing hardware upgrades. Various tools are accessible, including data collection from available features and \ac{ML} model training. The LLM can formulate a two-phase strategy: predict the traffic load and estimate the energy consumption for a specified load and hardware. It chooses the relevant \ac{ML} model for each subtask and indicates the data needed to train it. After devising the strategy, the LLM orchestrates available tools to collect the relevant data and train the \ac{ML} models.
}

\subsection{\acp{LLM} Limitations}
\label{subsec:llmlimitations}
\begin{figure}
    \centering
    \includegraphics[width=0.48\textwidth]
    {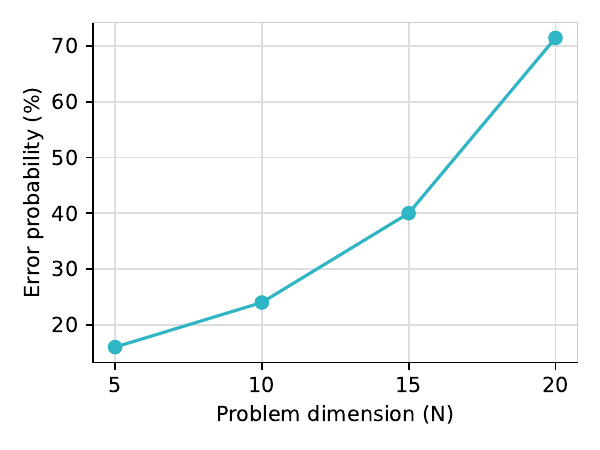}
    \caption{
    An illustration of an \ac{LLM} output inconsistency. GPT-3.5 was provided a vector reporting the strengths of N beams. It was tasked with selecting the beam with the highest strength, while instructed to avoid a particular beam, which was always set as the strongest. 
    }
    \label{fig:examplemisalignment}
\end{figure}

Given the structure and functionalities of \acp{LLM}, certain limitations become apparent. It is crucial to shed light on these shortcomings to utilize and interpret content generated by \acp{LLM}. The following are noteworthy flaws associated with them:
\subsubsection{Hallucinations and Fabrications} One of the key concerns with \acp{LLM} is their tendency to generate hallucinations or fabrications. \acp{LLM} rely on statistical patterns and associations learned from vast text data during training. Consequently, they may produce responses that abide to these patterns, but are incorrect or nonexistent \cite{bang2023multitask}.
   \subsubsection{Limited Explainability} The complex architecture and massive number of parameters in these models render it difficult to trace the decision-making process. In fact, \acp{LLM} lack transparency in terms of the specific features or patterns they rely on to generate responses. This opacity hinders the ability to understand why a particular answer or response was chosen over others. This limited explainability raises concerns, especially in domains where transparency and accountability are crucial. 
    \subsubsection{Computational Complexity} \acp{LLM} may consist of millions or even billions of parameters, making them resource-intensive to train and deploy. Even after training, running inference with \acp{LLM} can be computationally demanding. Generating responses with these models involves complex computations across multiple layers, which can strain available resources, especially for real-time applications. 
    \subsubsection{Sensitivity to Updates} \acp{LLM} display sensitivity to adjustments in their parameters, leading to unforeseen variations in outputs and behaviors. A compelling illustration of this phenomenon can be found in \cite{chen2023chatgpts}, which showcased how the performance and behavior of both GPT-3.5 and GPT-4 underwent dramatic shifts over time: in March 2023, GPT-4 excelled at identifying prime numbers, but by June 2023, it faltered in handling the same questions. This inconsistency serves as a clear illustration of the susceptibility of \acp{LLM} to updates and alterations introduced to the model.
    \subsubsection{Output Inconsistency} 
    This is a phenomenon that arises when the output generated by the model fails to fully align with the user's intent or the desired task, even when the prompt explicitly specifies the required output \cite{wolf2023fundamental}. This is illustrated in Fig.~\ref{fig:examplemisalignment}, where GPT-3.5 was tested to answer a simple constrained maximization question. The \ac{LLM} provided a wrong response with a given probability. Importantly, the error probability was observed to increase with the dimension of the problem.
    This can hamper the applicability of \acp{LLM} in areas such as telecom system optimization. Therefore, addressing this limitation becomes of utmost importance.
        
\section{\textcolor{black}{Potential LLM Applications In The Telecom Industry}}
\label{sec:LLM_Applications}
With the understanding of how \acp{LLM} function, their capabilities, and their limitations, we can now delve into the applications that can have a large impact on the telecom industry.
\subsection{Network Anomalies Resolution}
\begin{figure}
    \centering
    \includegraphics[width=0.4\textwidth]{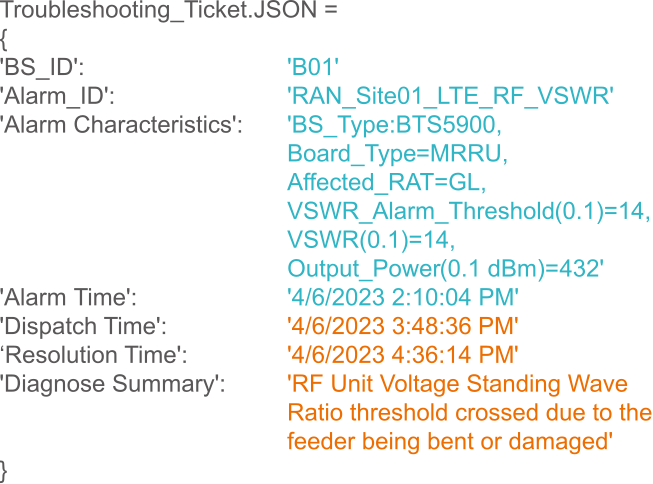}
    \caption{An example of a network anomaly troubleshooting ticket. Information related to the anomaly is automatically generated by the system (in blue). Input regarding the dispatch and the resolution of the anomaly is provided by the engineer (in orange).}
    \label{fig:troublesh}
\end{figure}
Solving anomalies in the mobile network is a tedious task. With a vast infrastructure spanning across large geographical areas, maintaining and monitoring the \acp{BS} is challenging. Each \ac{BS} is susceptible to a wide array of issues, including hardware malfunctions, software glitches, and environmental factors. For this reason, rectifying these anomalies necessitates extensive expertise, as arriving at appropriate solutions demands significant investments of manpower, meticulous analysis, and troubleshooting efforts. Leveraging \acp{LLM} can enhance the capabilities of \acp{MNO} in addressing these challenges and enable more efficient troubleshooting. Particularly, \acp{MNO} have at their disposal a rich repository of tickets accumulated over time from dealing with network anomalies. These tickets capture real-world scenarios, encompassing diverse problems and equipment malfunctions. An illustrative example of such a ticket 
is shown in Fig. \ref{fig:troublesh}. By utilizing this repository with product manuals as training data, the \ac{LLM} can be fine-tuned to comprehend the intricacies of network issues and grasp the unique context of anomaly resolution. Consequently, the \ac{LLM} becomes an anomaly-solving tool for telecommunications professionals, furnishing them with diagnoses of network issues and their corresponding solutions. Furthermore, leveraging the time-stamped data from the tickets, the \ac{LLM} can estimate the duration required to address network faults, accounting for the product type, hardware specificities, and the attributes of the involved \acp{BS}. 
As a result, the \ac{LLM} becomes an asset for the \ac{MNO}, enhancing the efficiency and effectiveness of resolving network problems.
\subsection{\ac{3GPP} Specifications Comprehension}
\textcolor{black}{The \ac{3GPP} produces the specifications that define cellular telecommunications
technologies, including radio access, core network and service capabilities.}
\ac{3GPP} documents are known for their elaborateness, encompassing many details and specifications. Due to the sheer volume of these documents, keeping track of all the specificities, especially in the context of new releases, can be daunting and time-consuming. For engineers attempting to implement technologies and features in the product, this challenge becomes even more apparent, as they must invest considerable time in searching for relevant information within the extensive documentation. \acp{LLM} offer a resolution to this problem, providing promising solutions for engineers grappling with \ac{3GPP} documents. Through fine-tuning to the \ac{3GPP} documents and incorporating all relevant reports, these models can become adept at processing the vast \ac{3GPP} standard knowledge. Then, a significant application of these fine-tuned \acp{LLM} revolves around the development of interactive chatbots tailored for answering \ac{3GPP} standards queries. These chatbots, built upon the fine-tuned \acp{LLM}, empower engineers to streamline their research processes, saving valuable time and facilitating more efficient and accurate implementations of \ac{3GPP} standards. 


\subsection{Network Modeling}


The optimization of mobile networks is a complex task that requires multiple models for capturing different \acp{KPI} of the network and the interactions between various network configuration parameters. Such optimization often relies on white-box models, where interactions between multiple features are mathematically formulated to ensure explainability. Developing such models requires expert engineers with deep domain knowledge to identify relevant information and relationships driving the interactions between features. Leveraging \acp{LLM} can support the development of these models.

To better clarify this aspect, we provide an explanatory example. Let us consider a simple scenario with a network composed of 90 single-carrier \acp{BS}. We used \mbox{GPT-3.5} as the \ac{LLM}. The \ac{LLM} was provided with a list of 12 data features, such as \ac{BS} location, frequency, and load, and tasked to select the relevant features for creating a model to estimate energy consumption based on the selected features. Additionally, we asked the \ac{LLM} to provide a mathematical formula capturing the relationship between inputs and outputs and a script to fit the model on a dataset containing real network data.
\textcolor{black}{\mbox{GPT-3.5}} successfully identified the 5 relevant inputs among the provided features, while discarding the irrelevant ones. Notably, this was achieved solely on the basis of its knowledge, without using any data samples. The model provided by \textcolor{black}{GPT-3.5} consisted of a weighted sum of the selected features for regression.

\begin{figure}
     \centering
     \subfloat[][\textcolor{black}{GPT-3.5}]{\includegraphics[scale=0.65]{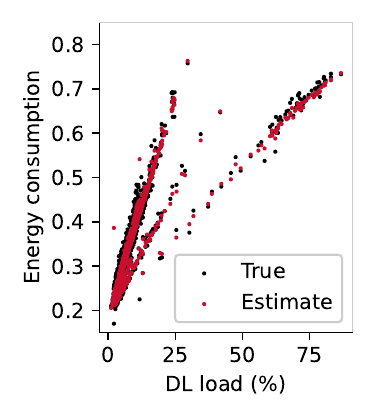}\label{<figure1>}}
     \subfloat[][\textcolor{black}{GPT-3.5} with context]{\includegraphics[scale=0.65]{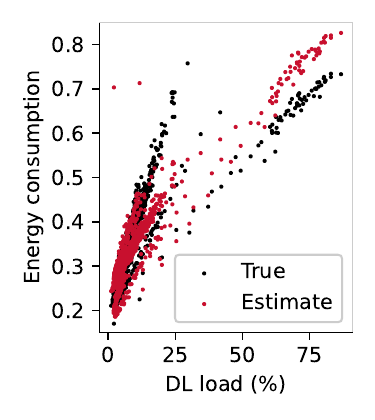}\label{<figure2>}}
     \caption{Normalized energy consumption measured at different downlink loads and estimated by the model provided by~(a) \textcolor{black}{GPT-3.5}, and~(b) \textcolor{black}{GPT-3.5} with context.}
     \label{fig:loadvsenergy}
\end{figure}

Fig.~\ref{fig:loadvsenergy} shows the real energy consumption measured by the \acp{BS} at different downlink loads. The real data reveal three different trends, corresponding to the three different configurations of maximum transmit powers in the considered network. 
Fig.~\ref{fig:loadvsenergy}a shows the estimations performed by the \textcolor{black}{model provided by GPT-3.5}, which achieved a relative error of 7.8\%. The estimations produced by this model resulted in a single average trend, as the selected inputs were summed, overlooking the relationship between the load and the maximum transmit power: in fact, these two terms should be multiplied and not summed. To address this limitation, we provided contextual data related to the dynamics driving the energy consumption of a generic \ac{BS}. By leveraging this, \textcolor{black}{GPT-3.5} produced a different model where the two terms were multiplied instead of summed, significantly reducing the error to 3\%. The improved model correctly captures all three trends (Fig.~\ref{fig:loadvsenergy}b), highlighting the importance of providing telecom-related context. 

\begin{figure}
    \centering
    \includegraphics[scale=0.8]{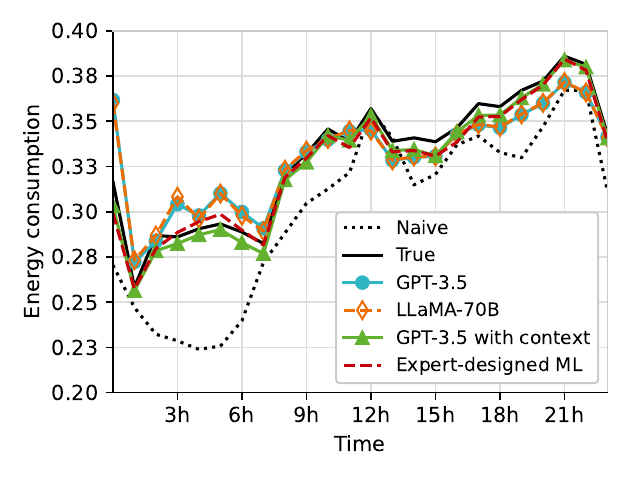}
    \caption{Normalized hourly energy consumption in the network - Actual measurements (in black) and estimations from various models.}
    \label{fig:weeklyenergy}
\end{figure}

Fig.~\ref{fig:weeklyenergy} illustrates the average hourly energy consumption in the selected network and the estimates performed by the two models provided by \textcolor{black}{GPT-3.5}  (i.e., with and without context).
To provide a basis for comparison, we present the estimations from two alternative models: i) a naive model and ii) an expert-designed \ac{ML} model~\cite{piovesan2022machine}.
The naive model estimates the energy consumption in a given hour by averaging the energy consumption measured at the same hour of the day in the previous week. While simple, this model lacks knowledge of the telecom field and consequently yields an error rate of 12\%.
On the other hand, the expert-designed \ac{ML} model employs a \ac{ML} model designed to handle more intricate scenarios, such as multi-carrier \acp{BS} utilizing multiple energy-saving features. In this simplistic setup, the expert-designed \ac{ML} model achieves a relative error of 2.3\%.
\textcolor{black}{Significantly, GPT-3.5 capitalized on its knowledge to develop a model that surpassed the limitations of the naive approach, realizing a 75\% improvement in accuracy, closely approaching the performance of the expert-designed \ac{ML} model.}

\textcolor{black}{As a final point, it is crucial to highlight that the choice of the LLM employed for a task significantly influences the quality of the achieved solution.}
\textcolor{black}{To illustrate this, we conducted the same experiment using LLaMA-70B as the \ac{LLM}. In this case, LLaMA identified additional input features compared to those selected by GPT-3.5, including location, and the year of production of the \ac{BS}. 
The model proposed by LLaMA took the form of a weighted sum of the chosen features, similar to the approach proposed by GPT-3.5, resulting in a similar error rate of 7.6\%. 
However, akin to GPT-3.5, the LLaMA model struggled to recognize the relationship between load and maximum transmit power. In contrast to GPT-3.5, though, LLaMA was unable to rectify this issue even when provided with additional contextual information.
}

\section{Open Research Directions}
\label{sec:researchdirections}
Extending upon the previously discussed limitations of \acp{LLM} and their forthcoming use cases in the telecom industry, a set of open research directions presents itself. These avenues of investigation are crucial to unlock the full potential of \acp{LLM} in the telecom industry and harness their capabilities to the utmost extent.
\subsection{Telecom Foundation Model}
\label{sec:Foundation Model}
{\color{black}
While the most advanced foundation models exhibit a reasonable grasp of the telecommunications theory, they fall short on practical implementation knowledge~\cite{maatouk2023teleqna}. 
Besides, our findings, illustrated in Fig.~\ref{fig:loadvsenergy}, have demonstrated the performance gap between a context-aware \ac{LLM} and a generic counterpart, shedding light on the necessity of a specialized telecom foundation model. This is further validated in~\cite{maatouk2023teleqna}. Such a specialized model should leverage standards, white papers, research literature, and even exclusive proprietary materials or synthetic datasets produced through simulators like digital twins. 

Three approaches are available to integrate further knowledge into a language model: full model training, fine-tuning, and \ac{RAG}. Full model training achieves a profound understanding of the additional knowledge at the expense of substantial energy and complexity costs. Fine-tuning offers a pragmatic balance, enabling model specialization via training a minimal number of parameters using methods like \ac{LoRA}. Meanwhile, \ac{RAG} is the most convenient solution. It is cost-efficient and does not require access to the model weights. It incorporates external knowledge using a more surface-level comprehension by querying a database for context to append to the prompt, which may limit the depth of understanding. 

It is worth mentioning that some lines of work investigate non-language-based telecom foundational models. Notably, graph-based foundational models could natively capture the natural topology of telecom networks. Such approaches remain in an early exploratory phase.
}

\subsection{\textcolor{black}{Benchmarking LLMs for Telecom}}
\textcolor{black}{In the last years researchers have proposed a number of tests to evaluate \ac{LLM} capabilities in terms of \ac{NLP}, e.g., text understanding and reasoning. Recent \acp{LLM} are already close to human-level performance on several of these tests such that HellaSwag, a test of commonsense inference, and GLUE/SuperGLUE, which evaluate \ac{LLM} linguistic understanding. MMLU, instead, evaluates \acp{LLM}’ multitask accuracy and capabilities across a broad range of subjects, and show that top-performing \acp{LLM} have still significant room for improvement before achieving expert-level accuracy across specialized tasks. In all these tests, accuracy on
 multiple-choice questions is computed to provide a simple to determine and understand evaluation.
 Some researchers have suggested that the future of \ac{NLP} evaluation should focus on text generation: however, while some metrics exist for testing these capabilities such as BLEU and perplexity, text generation is notoriously difficult to assess and still lacks a standard evaluation methodology. Although ML researchers have mainly focusing on \ac{NLP} capabilities of \acp{LLM}, the 
success of LLMs is the telecom industry depends on benchmark datasets designed to assess their proficiency in this specific domain. 
These datasets are expected to play a pivotal role in determining the optimal architectural design for \acp{LLM} and guiding the pretraining procedure in the development of telecom foundational models. The framework in~\cite{maatouk2023teleqna} proposes a multiple-choice question dataset to simply evaluate the accuracy of telecom knowledge of \acp{LLM}; future works will need to extend this framework and allow the evaluation of \acp{LLM} across specialized telecom tasks such as those discussed in Sec. \ref{sec:LLM_Applications}.}

\subsection{\acp{LLM} Compression}
As highlighted in Section \ref{subsec:llmlimitations}, \acp{LLM} can be comprised of billions of parameters and require powerful devices to be trained and inferred. 
This limitation becomes relevant in critical scenarios where \acp{LLM} need to be deployed in edge devices with limited storage and computational capabilities.
As a result, it is imperative to address the substantial size of \acp{LLM} and develop \textcolor{black}{compression techniques~\cite{dettmers2023qlora}}, which can reduce their size while retaining their knowledge of the telecom domain. \textcolor{black}{Pruning, quantization, and knowledge distillation are the three most popular model compression techniques for \ac{DL} models. Today, researchers believe that quantization outperforms pruning in most of \ac{LLM} architectures. Then,
due to the large costs of training LLMs, post-training quantization, where weights and activation tensors are encoded with a low-level of precision, e.g., 8-bit or 4bit instead of 16-bit, is the main adopted scheme. Indeed most of the open source \acp{LLM} offer quantized versions of larger models. In addition, knowledge distillation is currently explored to develop compact \acp{LLM} that can run on devices with limited resources. To conclude, compression methods reduce memory and computational resource usage but can degrade \ac{LLM} performance, and thus accuracy pre- and post compression has to be evaluated to analyse pros and cons of the existing and future techniques.}



\subsection{Privacy Considerations}
Adapting \acp{LLM} to address specific telecom-related tasks may require the use of datasets containing sensitive user information. In light of this, it becomes imperative to implement measures to protect privacy when handling such data. Included among these measures are data anonymization and aggregation, effectively removing personally identifiable information to protect individual privacy. The incorporation of techniques such as differential privacy is essential to ensure that these models remain impervious to leaking sensitive information during queries. \textcolor{black}{Additionally, the development of smaller \acp{LLM} that can run on edge devices will further enhance the end user's privacy.}

\subsection{Behavior Alignment}
Solving the problem of output inconsistency is essential to enable the adoption of \acp{LLM} in the telecom industry, especially in accuracy-critical areas. {
\color{black}
It has been shown that grounding LLMs with use-case-specific external tools, such as querying external knowledge with RAG, reduces hallucinations \cite{shuster2021retrieval}. Besides, it is crucial to incorporate mechanisms and metrics to assess the model's prediction confidence. Such mechanisms enable the identification of uncertain cases, triggering additional verification from humans in the loop. In order to measure prediction confidence, methods include using the LLM's internal evaluation of the likelihood of the output, generating multiple responses to a single query to assess consistency, or using one \ac{LLM} to review and refine the output of another.
}
Additionally, rigorous testing of \acp{LLM} against adversarial inputs and scenarios can help to reveal vulnerabilities and guide the development of reliable models. Finally, understanding prompt engineering is necessary, given that well-designed queries and instructions play a crucial role in shaping the model's behavior and ensuring accurate outputs. 

\subsection{\acp{LLM} Explainability}
The need for explainability in \acp{LLM} within the telecom industry is paramount due to stakeholder concerns regarding trust and reliance on \ac{ML} outputs, especially considering their limitations previously discussed in Section \ref{subsec:llmlimitations}. The adoption of \acp{LLM} for critical operations require a clear understanding of how and why specific outputs are generated. This necessitates the incorporation of explainability techniques such as referencing, where \acp{LLM} can provide sources or justifications for their responses. Additionally, explicitly integrating explainability objectives into the training process is crucial for this purpose. 

\subsection{Real-time Context}
{\color{black}
By design, \acp{LLM} are trained offline on large corpora of data and, therefore, are not aware of new findings that may be accessible through search engines. Consequently, prompting these \acp{LLM} can lead to potentially outdated answers, especially considering that the telecom industry continuously evolves with releases of new technical specifications. One approach to address this issue is to enable \acp{LLM} to access external tools. For instance, allowing \acp{LLM} to access the internet through dedicated channels, as OpenAI has done with ChatGPT. However, this approach confines the quality of \ac{LLM} generation to the outcomes derived from search queries. A more fundamental strategy is to create data pipelines to gather new relevant telecom knowledge. This knowledge can then be utilized by either augmenting queries through \ac{RAG} approaches (e.g., \color{black} as done by Grok, the LLM developed by XAI, with tweets\color{black}) or by conducting additional training of the \ac{LLM} to refine its parametric knowledge. The latter approach introduces various research possibilities, such as identifying the optimal frequency for updating the \ac{LLM}'s parametric knowledge and developing efficient methodologies for model updates to integrate the new material.
}

\subsection{Sustainability and Environmental Impact} 
{
\color{black}
Given their large parameter count, \acp{LLM} pose a substantial environmental concern in terms of carbon footprint. To mitigate these challenges, prioritizing smaller and more efficient models (e.g., Phi-2 that compete with larger models) is recommended.
Furthermore, incorporating efficient implementations of attention mechanisms and overall model architecture can substantially alleviate computational demands during both training and inference. For instance, adopting the FlashAttention mechanism \cite{dao2022flashattention} or employing the mixture of experts architecture, as demonstrated by models like Mixtral, offers promising avenues for reducing computational loads. From another perspective, tackling the sustainability challenge also involves the development of KPIs and regulations that effectively measure, evaluate, and compare the environmental footprint of different \acp{LLM}.
}

\subsection{\color{black}LLMs as Orchestrators}
\textcolor{black}{Leveraging even further the reasoning capabilities of the LLMs, an open research direction involves transitioning from a strict parametric knowledge framework to a different paradigm where LLMs serve as orchestrators, as introduced in Section \ref{sec:demystifying}. In this scenario, LLMs are granted access to fine-grained blocks, such as code interpreters, optimizers, signal processing blocks, and network models. Their role then shifts to translating user prompts into actionable steps by leveraging both their knowledge and harnessing the accessible blocks. In this context, the research avenues revolve around defining these fine-grained blocks and ensuring seamless integration between LLMs and these blocks to unlock their potential.}

\section{Conclusions}
\label{sec:conclusions}
In this article, we have delved into the inner workings of LLMs, shedding light on their current capabilities and limitations. Additionally, we explored various use cases of LLMs that can be promptly leveraged within the industry using the available data at vendors' disposal. Furthermore, we discussed the specific open research directions tailored to the peculiarities of the telecom domain, which must be addressed to fully harness the potential of LLMs. As the technology behind LLMs continues to evolve, the telecom industry is poised to seize the opportunity and leverage these advancements to enhance operational efficiency within the sector.


\bibliographystyle{IEEEtran}

\bibliography{reference.bib}
\vspace{-1.1cm}
\begin{IEEEbiographynophoto}
    {Ali Maatouk} is a Researcher with Huawei Technologies, France. 
\end{IEEEbiographynophoto}
\vspace{-1.1cm}
\begin{IEEEbiographynophoto}
    {Nicola Piovesan} is a Senior Researcher with Huawei Technologies, France. 
\end{IEEEbiographynophoto}
\vspace{-1.1cm}
\begin{IEEEbiographynophoto}
    {Fadhel Ayed} is a Senior Researcher with Huawei Technologies, France. 
\end{IEEEbiographynophoto}
\vspace{-1.1cm}
\begin{IEEEbiographynophoto}
    {Antonio De Domenico}
is a Senior Researcher with Huawei Technologies, France. 
\end{IEEEbiographynophoto}
\vspace{-1.1cm}
\begin{IEEEbiographynophoto}
    {M\'erouane Debbah} is 
    a Professor at  Khalifa University in Abu Dhabi. 
\end{IEEEbiographynophoto}

\begin{acronym}[AAAAAAAAA]
  \acro{3GPP}{Third Generation Partnership Project}
 \acro{AI} {Artificial Intelligence}
 \acro{BERT}{Bidirectional Encoder Representations from Transformers}
  \acro{DL}{Deep Learning}
   \acro{FPGA}{Field-Programmable Gate Array}
  \acro{GPT}{Generative Pre-trained Transformer}
  \acro{KPI}{Key Performance Indicator}
 \acro{LLM}{Large Language Model}
\acro{BS}{Base Station}
  \acro{ML}{Machine Learning}
  \acro{MNO}{Mobile Network Operator}
 \acro{NLP}{Natural Language Processing} 
\acro{NN}{Neural Network}
 \acro{RAN}{Radio Access Network}
\acro{LoRA}{low-rank adaptation}
\acro{RAG}{Retrieval Augmented Generation}
\acro{API}{Application Programming Interface}
 \end{acronym}

\end{document}